\documentclass[twocolumn,amsmath,prd]{revtex4}
\include{psfig}

\def\et{\eta}
\def\th{\theta}

\def\ta{\tau}

\def\ph{\phi}

\def\ch{\chi}

\def\om{\omega}

\def\mn{{\mu\nu}}

\def\fr#1#2{{{#1} \over {#2}}}
\def\frac#1#2{\textstyle{{{#1} \over {#2}}}}
\def\pt#1{\phantom{#1}}

\def\vev#1{\langle {#1}\rangle}

\def\half{{\textstyle{1\over 2}}}
\def\lsim{\mathrel{\rlap{\lower4pt\hbox{\hskip1pt$\sim$}}
    \raise1pt\hbox{$<$}}}
\def\gsim{\mathrel{\rlap{\lower4pt\hbox{\hskip1pt$\sim$}}
    \raise1pt\hbox{$>$}}}

\def\etal {{\it et al.}}
\newcommand{\beq}{\begin{equation}}
\newcommand{\eeq}{\end{equation}}
\newcommand{\bea}{\begin{eqnarray}}
\newcommand{\eea}{\end{eqnarray}}
\newcommand{\bse}{\begin{subequations}}
\newcommand{\ese}{\end{subequations}}
\newcommand{\rf}[1]{(\ref{#1})}

\def\to{\rightarrow}

\def\mix{\leftrightarrow}

\def\aL{(a_L)}
\def\cL{(c_L)}
\def\nub{\bar\nu}

\def\heff{h_{\rm eff}}
\def\As#1{({\cal A}_s)_{#1}}
\def\Ac#1{({\cal A}_c)_{#1}}
\def\Bs#1{({\cal B}_s)_{#1}}
\def\Bc#1{({\cal B}_c)_{#1}}
\def\C#1{({\cal C})_{#1}}
\def\Asa#1{({\cal A}_s^{(0)})_{#1}}
\def\Asc#1{({\cal A}_s^{(1)})_{#1}}
\def\Aca#1{({\cal A}_c^{(0)})_{#1}}
\def\Acc#1{({\cal A}_c^{(1)})_{#1}}
\def\Bsc#1{({\cal B}_s^{(1)})_{#1}}
\def\Bcc#1{({\cal B}_c^{(1)})_{#1}}
\def\Ca#1{({\cal C}^{(0)})_{#1}}
\def\Cc#1{({\cal C}^{(1)})_{#1}}
\def\nh{{\hat N}}

\begin{document}
\title{Lorentz violation and short-baseline neutrino experiments}
\author{V.\ Alan Kosteleck\'y and Matthew Mewes}
\affiliation{Physics Department, Indiana University, 
         Bloomington, IN 47405, U.S.A.}
\date{IUHET 470, June 2004} 

\begin{abstract}
A general discussion is given of signals for broken Lorentz symmetry 
in short-baseline neutrino experiments.
Among the effects that Lorentz violation can introduce 
are a dependence on energy differing from 
that of the usual massive-neutrino solution
and a dependence on the direction of neutrino propagation.
Using the results of the LSND experiment,
explicit analysis of the effects of broken Lorentz symmetry 
yields a nonzero value $(3\pm 1)\times 10^{-19}$ GeV 
for a combination of coefficients for Lorentz violation.
This lies in the range expected for effects originating 
from the Planck scale in an underlying unified theory.
\end{abstract}
\maketitle


Evidence for neutrino oscillations from the
Liquid Scintillator Neutrino Detector (LSND) experiment 
remains one of the unresolved anomalies in neutrino physics
\cite{lsnd1,lsnd2}.
The results of this experiment are consistent with 
muon antineutrinos $\nub_\mu$ oscillating into 
electron antineutrinos $\nub_e$
with a small probability of 
$P_{\nub_\mu\to\nub_e}\simeq 0.26\pm 0.08\%$
as the neutrinos propagate over a distance $L\simeq 30$ m.
Typically, 
the oscillations are explained
by ascribing small masses to neutrinos.
The anomaly exists because the LSND result contradicts
other neutrino-oscillation experiments 
in analyses based on three massive neutrinos
\cite{mv}.
More exotic oscillation scenarios may provide 
reconciliation of this contradiction.
In the present work,
we consider the possibility that the anomalous oscillations 
are induced by violation of Lorentz symmetry.
We show that the size of the violation
required to account for the anomaly 
is consistent with the size of effects that would emerge 
from an underlying unified theory at the Planck scale.

The principle of Lorentz symmetry is deeply ingrained in 
our best theories of nature at the fundamental level,
which are Einstein's general relativity
and the Standard Model of particle physics.
The former accounts for gravitation at the classical level,
while the latter describes all other physical phenomena
involving elementary particles and forces down to the quantum level.
However,
a unification of these two theories is expected to occur
at the Planck scale,
$m_P \simeq 10^{19}$ GeV,
about 17 orders of magnitude above the electroweak scale.
Among the candidate experimental signals of new physics
arising from this underlying unified theory
are minuscule violations of Lorentz symmetry
\cite{kps,cpt01}.

Despite an expected suppression 
by $10^{-17}$ or more,
Lorentz violation may be detectable with existing technology
in experiments on a variety of systems,
including neutrinos
\cite{ck,cg,bpww,bbm,mp,dg,kmnu,gl,chki,dgms}.
Under favorable circumstances,
attainable sensitivities in the neutrino sector
are competitive with those in other Lorentz tests,
including ones with
mesons \cite{hadronexpt},
baryons \cite{ccexpt},
electrons \cite{eexpt,eexpt2,eexpt3},
photons \cite{photonexpt,photonth,cavexpt,kmphot},
muons \cite{muons},
and the Higgs
\cite{higgs}.
Indeed,
the current evidence for neutrino oscillations
is compatible with small Lorentz violation,
perhaps even without the introduction of neutrino mass
\cite{kmnu}.
The novel features introduced by Lorentz violation 
may also make possible a reconciliation of the LSND result 
with other neutrino experiments.
The confirmation of oscillations from Lorentz violation 
in LSND and other neutrino experiments
would offer the first glimpse of nature at its most basic level.

General Lorentz violation is described by a theoretical framework 
known as the Standard-Model Extension (SME) 
\cite{ck,akgrav}.
It connects to the Planck scale
through operators of nonrenormalizable dimension
\cite{kple}.
In the minimal form of the SME,
neutrinos are massless and the oscillations 
are determined by a set of 102 real constant coefficients 
controlling the Lorentz violation
\cite{kmnu}.
These coefficients can be collected into matrices 
$(a_L)^\mu$ and $(c_L)^{\mu\nu}$
of mass dimension 1 and 0, respectively,
corresponding to CPT-odd and CPT-even operators.
Setting all the coefficients to zero yields
the usual minimal Standard Model of particle physics.
Assuming quantum-gravity origins for
$(a_L)^\mu$ and $(c_L)^{\mu\nu}$
suggests their magnitude is suppressed 
by a factor of $10^{-17}$ or more.

The coefficients for Lorentz violation predict 
a plethora of unconventional signals 
in oscillation experiments.
For example,
the usual mass-induced oscillations vary as $L/E$
with the neutrino energy $E$ and the experimental baseline $L$.
In contrast,
the simplest Lorentz-violating oscillations vary as $L$ or $EL$,
and more complicated functional dependence on $E$ can occur. 
This unconventional energy dependence can help reconcile 
conflicts with other experiments.
For instance, 
within the massive-neutrino scenario
the evidence for oscillations in LSND
implies that the lower-energy CHOOZ experiment
\cite{chooz}
should also have observed oscillations,
contrary to the experimental results.
However, 
the SME contains coefficients that predict oscillations 
in LSND without significant oscillations in CHOOZ 
\cite{kmnu}.

Another unconventional feature of oscillations due to Lorentz violation
is dependence on the direction of neutrino propagation.
This causes several novel effects,
which can include varying oscillation signals as the Earth rotates. 
For experiments like LSND, 
where both source and detector are fixed on the Earth's surface,
the entire apparatus makes a full rotation each sidereal day
$\simeq23$ h 56 min.
The direction of propagation of the detected neutrinos 
is therefore also changing,
and in the presence of Lorentz violation
the consequent change in oscillation probabilities 
provides a definite signal 
that cannot arise from mass-induced oscillations
\cite{kmnu}.
This effect provides another path 
to reconciling the apparent conflict of LSND with other experiments.
Thus,
for example, 
LSND and the KARMEN experiment 
\cite{karmen}
conflict in the massive-neutrino case.
However,
the neutrinos in these two experiments
propagate in different directions,
so they can indeed behave differently 
in Lorentz-violating scenarios.

A more exotic possibility,
absent in the minimal form of the SME 
but allowed in the general SME framework,
is oscillations between neutrinos and antineutrinos
\cite{kmnu}.
Oscillations with $\nu\mix\nub$ mixing 
offer alternative Lorentz-violating modes 
that could explain the excess of $\nub_e$ 
observed by LSND,
since the numbers of $\nu_e$, $\nu_\mu$,
and $\nub_\mu$ involved are comparable.
In what follows,
we restrict attention to the minimal SME scenario.

The large number of coefficients involved,
even in the minimal SME,
makes a general analysis challenging.
However, 
in experiments like LSND,
the short baseline offers the possibility 
of a valuable simplifying approximation.
When the baseline is small compared to the oscillation
lengths given by the hamiltonian,
the transition amplitudes can be expanded about the identity
as a perturbation on the oscillation-free case.
It turns out that the general leading-order result 
for the corresponding transition probabilities 
differs from the oscillation-free case 
by terms proportional to the squared modulus
of hamiltonian elements,
as we show next.

In the minimal Standard-Model Extension,
the oscillatory behavior of the three generations
of left-handed neutrinos 
is governed by the leading-order effective hamiltonian
\cite{kmnu}
\beq
\heff=
\fr{1}{E}\big[\aL^\mu p_\mu-\cL^\mn p_\mu p_\nu\big] .
\label{heff}
\eeq
In this equation,
$\aL^\mu$ and $\cL^\mn$ are coefficients for Lorentz violation
that are hermitian $3\times3$ complex matrices
of mass dimension 1 and 0, respectively.
The energy $E$ is assumed to be large compared 
to the elements of $\aL^\mu$ and $E\cL^\mn$.
The four-momentum $p_\mu\simeq E(1;-\hat p)$
introduces both energy dependence through $E$
and direction dependence through $\hat p$.
Since the antisymmetric and trace pieces of $\cL^\mn$
do not contribute to Eq.\ \rf{heff},
we also assume in what follows the properties 
$\cL^\mn=\cL^{\nu\mu}$ and $\et_\mn\cL^\mn=0$.
The effective hamiltonian for antineutrinos
is obtained by complex conjugating Eq.\ \rf{heff}
and reversing the sign of the $\aL^\mu$ term.

Under suitable experimental conditions,
it is an excellent approximation to expand 
the oscillation amplitudes in powers of $\heff$: 
$S(L)\simeq 1-i\heff L/(\hbar c)
-\half\heff^2 L^2/(\hbar c)^2+\cdots$.
The validity of this expansion
requires that the baseline $L$ be short compared 
to the oscillation lengths given by $\heff$.
However,
since $\heff$ varies with the neutrino energy $E$, 
the designation of a given experiment as short baseline 
in this context depends on the ranges of both $L$ and $E$.
At leading order in this short-baseline approximation,
the oscillation probabilities are 
\beq
P_{\nu_b\to\nu_a}\simeq
\left\{
  \begin{array}{lr}
    1-\sum_{c, c\ne a}
    P_{\nu_a\to\nu_c}, \quad & a=b\ ,\\[10pt]
    |(\heff)_{ab}|^2L^2/(\hbar c)^2, & a\ne b\ ,
  \end{array}
  \right.
  \label{P}
\eeq
where the indices $a,b$ range over the neutrino flavors
$e, \mu, \ta$.
The probabilities $P_{\nub_b\to\nub_a}$ for antineutrinos
are obtained by changing the sign of $\aL^\mu$.
Note that Eq.\ \rf{P}
can readily be modified for the nonminimal SME, 
including $\nu\mix\nub$ mixing
\cite{kmnu}.

In reporting results from experimental tests of Lorentz invariance,
it is necessary to specify the frame of reference.
In principle,
any inertial frame can be adopted,
but convention and convenience dictate the use 
of a Sun-centered celestial-equatorial frame.
For experiments with both source and detector 
fixed on the Earth's surface,
the sidereal rotation causes 
the direction of neutrino propagation $\hat p$
to change with respect to the Sun-centered frame.
This causes the components of $\hat p$ 
to vary at the sidereal frequency
$\om_\oplus=2\pi/(23$ h 56 min),
unless $\hat p$ happens to point along the Earth's rotation axis.
This time dependence can be displayed explicitly 
in the effective hamiltonian $\heff$,
which can be written in the form 
\def\indx{{ab}}
\begin{align}
  (\heff)_\indx &=
  \C\indx
  +\As\indx
  \sin\om_\oplus T_\oplus
  +\Ac\indx
  \cos\om_\oplus T_\oplus
  \nonumber \\ &\quad
  +\Bs\indx
  \sin2\om_\oplus T_\oplus
  +\Bc\indx
  \cos2\om_\oplus T_\oplus ,
  \label{heff2}
\end{align}
where $T_\oplus$ is the time measured from a standard origin
\cite{kmphot}.
This expression is independent of the short-baseline approximation,
so Eq.\ \rf{heff2} and what follows  
also apply more generally.

The energy dependence in Eq.\ \rf{heff2}
is given by further decomposition:
\begin{align}
  \C\indx
  &= \Ca\indx+E\Cc\indx ,
  \nonumber\\
  \As\indx
  &= \Asa\indx+E\Asc\indx ,
  \nonumber\\
  \Ac\indx
  &= \Aca\indx+E\Acc\indx ,
  \nonumber\\
  \Bs\indx
  &= E\Bsc\indx ,
  \quad
  \Bc\indx
  = E\Bcc\indx .
\label{decomp}
\end{align}
The combinations
$\Asa\indx$,
$\Aca\indx$,
$\Ca\indx$
contain the coefficients $\aL^\mu$, 
while
$\Asc\indx$,
$\Acc\indx$,
$\Bsc\indx$,
$\Bcc\indx$,
$\Cc\indx$
involve the coefficients $\cL^\mn$.
The analogous decomposition 
for the antineutrino effective hamiltonian 
generates combinations 
that can be obtained from their neutrino equivalents 
by complex conjugation and a sign reversal for 
$\Asa\indx$,
$\Aca\indx$, 
$\Ca\indx$.

The explicit relationships between these quantities 
and the SME coefficients 
$\aL^\mu$ and $\cL^\mn$ for Lorentz violation are 
\begin{align}
  \Ca\indx &=
  \aL_\indx^T-\nh^Z\aL_\indx^Z
  ,\\
  \Cc\indx &=
  -\half(3-\nh^Z\nh^Z)\cL_\indx^{TT}
  +2\nh^Z\cL_\indx^{TZ}
  \nonumber \\
  &\quad+\half(1-3\nh^Z\nh^Z)\cL_\indx^{ZZ} ,\\
  \Asa\indx &=
  \nh^Y\aL_\indx^X-\nh^X\aL_\indx^Y ,\\
  \Asc\indx &=
  -2\nh^Y\cL_\indx^{TX}+2\nh^X\cL_\indx^{TY}
  \nonumber \\
  &\quad+2\nh^Y\nh^Z\cL_\indx^{XZ}-2\nh^X\nh^Z\cL_\indx^{YZ} ,\\
  \Aca\indx &=
  -\nh^X\aL_\indx^X-\nh^Y\aL_\indx^Y ,\\
  \Acc\indx &= 
  2\nh^X\cL_\indx^{TX}+2\nh^Y\cL_\indx^{TY}
  \nonumber \\
  &\quad-2\nh^X\nh^Z\cL_\indx^{XZ}-2\nh^Y\nh^Z\cL_\indx^{YZ} ,
\end{align}
\begin{align}
  \Bsc\indx &=
  \nh^X\nh^Y
  \big(\cL_\indx^{XX}-\cL_\indx^{YY}\big)
  \nonumber \\
  &\quad-\big(\nh^X\nh^X-\nh^Y\nh^Y\big)
  \cL_\indx^{XY} ,\\
  \Bcc\indx &=
  -\half\big(\nh^X\nh^X-\nh^Y\nh^Y\big)
  \big(\cL_\indx^{XX}-\cL_\indx^{YY}\big)  
  \nonumber \\
  &\quad-2\nh^X\nh^Y\cL_\indx^{XY} .
\end{align}
In these expressions,
$\nh^X$, $\nh^Y$, $\nh^Z$ are directional factors 
containing information about the neutrino-beam direction
with respect to the Earth.
At the detector location,
let $\th$ be 
the angle between the beam and the vertical upward direction,
let $\ph$ be 
the angle between the beam and south measured towards the east,
and let $\ch$ be the colatitude of the detector.
Then,
the directional factors are given explicitly as
\beq
\left(\begin{array}{c}
    \nh^X\\ \nh^Y\\ \nh^Z
  \end{array}\right)
=\left(\begin{array}{c}
    \cos\ch\sin\th\cos\ph+\sin\ch\cos\th \\
    \sin\th\sin\ph \\
    -\sin\ch\sin\th\cos\ph+\cos\ch\cos\th
  \end{array}\right) .
\eeq

Any given short-baseline experiment is sensitive 
to three complex combinations of $\aL^\mu$ coefficients,
$\Asa\indx$,
$\Aca\indx$, 
$\Ca\indx$,
and five complex combinations of
$\cL^\mn$coefficients,
$\Asc\indx$, 
$\Acc\indx$,
$\Bsc\indx$,
$\Bcc\indx$,
$\Cc\indx$.
However,
the directional dependence implies that
a combination of experiments testing a specific oscillation mode
$\nu_a\to\nu_b$ 
can provide access to all components of
$\aL^\mu_\indx$ and
$\cL^\mn_\indx$,
provided the directions of the associated neutrino beams differ.

For the special case of the transition mode 
relevant to LSND,
the probability takes the form
\def\indx{{\bar e\bar\mu}}
\begin{align}
  &P_{\nub_\mu\to\nub_e}\simeq
  \fr{L^2}{(\hbar c)^2} |\, \C\indx
  \nonumber \\ &\quad
\pt{xxxxxxxx}
  +\As\indx
  \sin\om_\oplus T_\oplus
  +\Ac\indx
  \cos\om_\oplus T_\oplus
  \nonumber \\ &\quad
\pt{xxxxxxxx}
  +\Bs\indx
  \sin2\om_\oplus T_\oplus
  +\Bc\indx
  \cos2\om_\oplus T_\oplus \, |^2,
\label{lsnd}
\end{align}
where $\om_\oplus\simeq2\pi/(23$ h 56 min)
is the Earth's sidereal frequency
and $T_\oplus$ is a standardized time 
\cite{kmphot}.
The time variation is a direct consequence 
of the directional dependence.
In the short-baseline approximation,
we find harmonics up to $2\om_\oplus$,
but more generally all higher harmonics can occur. 

In Eq.\ \rf{lsnd},
the complex factors
$\As\indx$, $\Ac\indx$,
$\Bs\indx$, $\Bc\indx$,
and $\C\indx$
are experiment-dependent linear combinations
of the SME coefficients $(a_L)^\mu$ and $(c_L)^{\mu\nu}$
for Lorentz violation.
These combinations depend on the energy of the neutrinos.
Their decomposition into energy-independent quantities 
takes a form analogous to that of Eq.\ \rf{decomp}:
\begin{align}
  \C\indx &= \Ca\indx+E\Cc\indx ,
  \nonumber\\
  \As\indx
  &= \Asa\indx+E\Asc\indx ,
  \nonumber\\
  \Ac\indx
  &= \Aca\indx+E\Acc\indx ,
  \nonumber\\
  \Bs\indx
  &= E\Bsc\indx , \quad \Bc\indx = E\Bcc\indx .
\label{decomp2}
\end{align}
There are therefore a total of eight complex 
experiment-dependent coefficients:
$\Asa\indx$,
$\Aca\indx$, 
$\Ca\indx$,
$\Asc\indx$, 
$\Acc\indx$,
$\Bsc\indx$,
$\Bcc\indx$,
$\Cc\indx$.
A comprehensive analysis of the LSND data 
for the above energy and sidereal dependence
would in principle yield measurements of 
16 of the possible 102 real degrees of freedom 
in the neutrino sector of the minimal SME.
We remark in passing that the inclusion of a mass-squared matrix
$(\tilde m^2)_{ab}$ for neutrinos
in the present formalism is straightforward.
For example, 
in Eq.\ \rf{decomp2}
it suffices to extend the definition of $\C\indx$ 
to $\C\indx = (2E)^{-1} (\tilde m^2)^*_\indx + \Ca\indx+E\Cc\indx$.
It turns out that
the general two-generation model with a mass-squared matrix
and both $\aL^\mu$ and $\cL^\mn$ coefficients 
has 41 degrees of freedom,
while its rotation-invariant restriction has eight
\cite{kmnu}.

\begin{figure}
\centerline{\psfig{figure=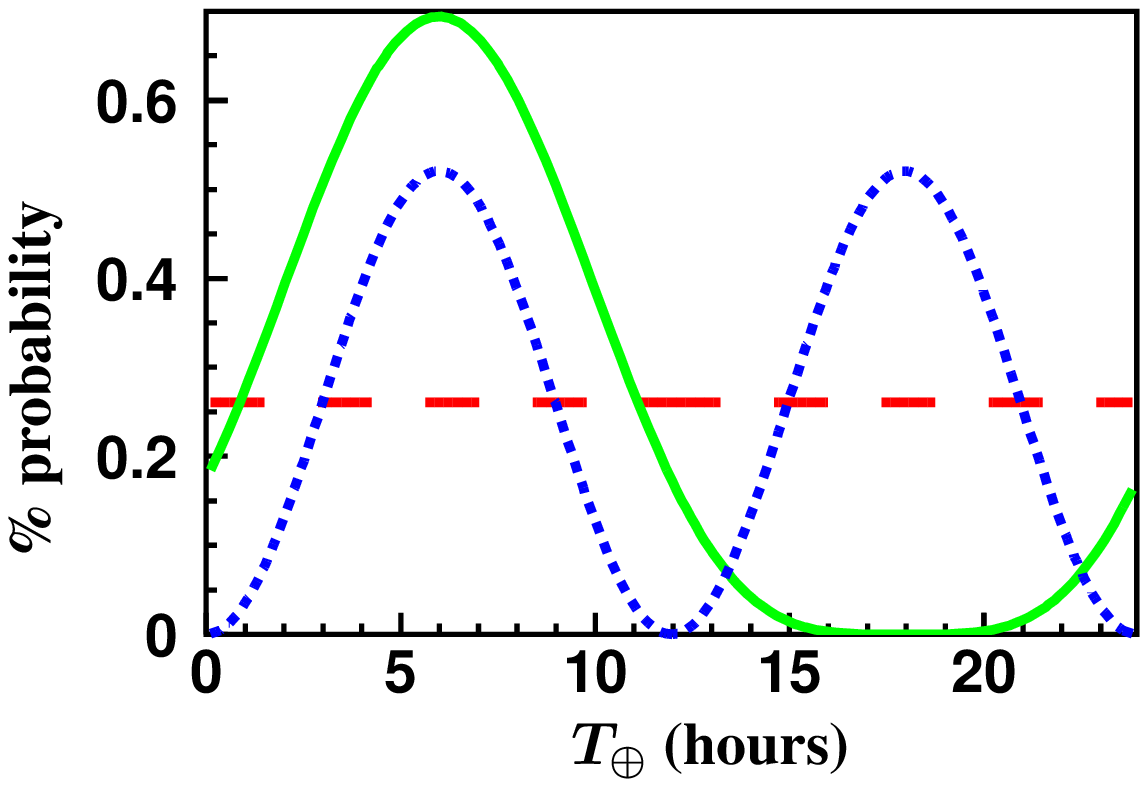,width=0.8\hsize}}
\caption{\label{fig1}
Variations of the percent probability $P_{\nub_\mu\to\nub_e}$
over one sidereal day for three sample configurations with
averaged probability $\vev{P_{\nub_\mu\to\nub_e}}=0.26\%$:
$\C\indx\ne 0$ (dashed),
$\As\indx\ne 0$ (dotted),
and $\C\indx=\As\indx\ne 0$ (solid).
}
\end{figure}

The published results from LSND permit the extraction of a measurement
for one combination of these degrees of freedom.
In the experiment,
copious numbers of $\nub_\mu$ were produced.
An excess of $\nub_e$ over background was observed,
which was interpreted as $\nub_\mu$ oscillating into $\nub_e$.
The corresponding oscillation probability is 
$P_{\nub_\mu\to\nub_e}\simeq 0.26\pm 0.08\%$.
Since this published result involves all events 
irrespective of sidereal time,
it represents an average over the run time of the experiment.
To a good approximation,
it can be taken as representing the expectation 
over a sidereal day,
$\vev{P_{\nub_\mu\to\nub_e}}\simeq 0.26\pm 0.08\%$.
Using Eq.\ \rf{lsnd} and this result,
we obtain a nonzero measurement
for a combination of SME coefficients for Lorentz violation:
\begin{align}
  &|\C\indx|^2+\half|\As\indx|^2+\half|\Ac\indx|^2
  \nonumber \\
  &\qquad+\half|\Bs\indx|^2+\half|\Bc\indx|^2
  \nonumber \\
  &\quad\simeq \fr{(\hbar c)^2 \vev{P_{\nub_\mu\to\nub_e}}}{L^2}
  \nonumber \\
  &\quad\simeq \big((3\pm 1)\times10^{-19}\mbox{ GeV }\big)^2 .
  \label{meas}
\end{align}
Since the LSND neutrino energy 
lies in the range 10 MeV $\lsim E \lsim$ 50 MeV,
this result corresponds to values of the 
SME coefficients for Lorentz violation 
of order $10^{-19}$ GeV for $(a_L)^\mu$ 
and $10^{-17}$ for $(c_L)^{\mu\nu}$.
Remarkably, 
these values are in the range predicted for quantum-gravity effects.

Establishing which specific combinations 
of the 16 possible degrees of freedom
could be predominantly responsible 
for the LSND signal is more challenging 
and requires information about sidereal and energy dependences.
Figure \ref{fig1} illustrates some of the various possibilities.
The probability $P_{\nub_\mu\to\nub_e}$ is displayed as a function 
of sidereal time for three situations with distinct combinations 
of nonzero coefficients $\C\indx$ and $\As\indx$.
The probabilities differ in detail, 
but all yield the result \rf{meas}.

Other short-baseline experiments 
\cite{chooz,karmen,chorus,nomad,nutev,miniboone}
could perform similar analyses 
to obtain further information about the space 
of coefficients for Lorentz violation.
If the above solution is confirmed,
it would constitute a convincing signal for Lorentz violation
and could offer the first experimental glimpse of 
Planck-scale physics.

This work was supported in part
by D.O.E. grant DE-FG02-91ER40661
and N.A.S.A. grants NAG8-1770 and NAG3-2194.

\end{document}